# Interatomic force laws that corrupt their own measurement

John E. Sader[1,2,*] Barry D. Hughes[2], Ferdinand Huber[3], and Franz J. Giessibl[3]

[1] *ARC Centre of Excellence in Exciton Science, The University of Melbourne, Victoria 3010, Australia*

[2] *School of Mathematics and Statistics, The University of Melbourne, Victoria 3010, Australia*

[3] *Institute of Experimental and Applied Physics, University of Regensburg, D-93053 Regensburg, Germany*

## Abstract

Atomically-resolved imaging and force measurements using the atomic force microscope (AFM) are performed most commonly in a frequency-modulation (FM) mode. This has led to spectacular results, including direct observation of the atomic structure of complex molecules and quantification of chemical and frictional forces at the atomic scale. We address here a critical question: Is recovery of force from the measured frequency shift experienced by the AFM cantilever *ill-posed*—that is, unreliable in the presence of (unavoidable) measurement uncertainty? Resolution of this issue underlies all force measurements using FM–AFM, but remains outstanding. It is shown that concavity of the force law's distance dependence controls ill-posed behavior, with a rapid concavity change corrupting force measurements by inducing spurious and unphysical effects—such rapid change is not uncommon. Practical conditions to eliminate ill-posed behavior are formulated which are verified experimentally. This study lays the foundations for robust atomically-resolved force spectroscopy and future work that will seek to regularize ill-posed force measurements.

The foundation of all mechanical sensors is their ability to transduce an applied signal, such as force or added mass, to a measurable change in the sensor's response. This response is often analyzed using mathematical models that enable quantitative information about the applied signal to be recovered, e.g., the inertial mass of an adsorbate from the measured change in resonant frequency of the sensor [1–5]. Fundamental to these applications is the overriding assumption that (small) measurement uncertainty does not render the signal inversion procedure/algorithm unreliable—that is, the inversion process is not *ill-posed*.

The success of the atomic force microscope (AFM) in quantifying structure and forces at the molecular and atomic scale hinges on the extreme precision with which the response of its force sensing microcantilever can be measured [6, 7]. Changes in the AFM cantilever's deflection at the picometer scale, and absolute measurement of its resonant frequency at the ppm level are performed routinely. These observables can be converted into force that the cantilever experiences through a range of established mathematical algorithms [8–10]. This has led to tremendous advances including the quantification and direct measurement of friction and chemical forces at the atomic scale [11–19].

Static measurements of force that monitor the cantilever's deflection are obviously not ill-posed, i.e., they are *well-posed*, because the recovered force is linearly proportional to the measured deflection, via Hooke's law. However, the situation is not clear for dynamic measurements, where the cantilever's response depends on the force applied to its tip in a more complex manner. In dynamic measurements, the cantilever often experiences a range of forces as it oscillates dynamically, leading to a convoluted relationship between the interaction force and the measured resonant frequency, amplitude and/or phase of the cantilever's motion [6].

In frequency-modulation (FM) AFM [20], which is widely used for atomically-resolved measurements, the cantilever is self-excited in a feedback loop that guarantees it oscillates on resonance. A second feedback loop monitors and controls its oscillation amplitude. In this way, minute changes in the cantilever's resonant frequency at fixed oscillation amplitude can be detected in (quasi) real-time. While providing marked improvements in sensitivity over static measurements, the FM–AFM methodology for determining force from the measured frequency shift and oscillation amplitude is more complex. Such FM *force spectroscopy* measurements require solution to the integral equation [21]

$$\Omega(z) = -\frac{1}{\pi a k}\int_{-1}^{1} F(z + a(1+u))\frac{u}{\sqrt{1-u^2}}\, du, \tag{1}$$

where $\Omega(z) \equiv \Delta\omega(z)/\omega_{\text{res}}$ is the measured relative change in the cantilever's resonant frequency resulting from the (to be determined) interaction force, $F(z)$, between the cantilever tip and sample, $\omega_{\text{res}}$ is the unperturbed resonant frequency of the cantilever, $\Delta\omega(z)$ is the change in resonant frequency, $k$ is the dynamic spring constant, $a$ is the oscillation amplitude, and $z$ ($\geq 0$) is the distance of closest approach between tip and sample. In practice, the origin of $z$ can be shifted such that $z = 0$ is the smallest distance from which data is measured. This convention is used here.

**The inversion problem**

A number of mathematical approaches have been developed to determine the interaction force by inversion of Eq. (1). Dürig [22] formulated a method to recover the force from the frequency shift in the limit of large oscillation amplitude, building on the original work of Albrecht *et al.* [20] in the small amplitude limit. For arbitrary amplitudes—the general practical case—perhaps the two most commonly used methods are those due to (i) Giessibl [9], and (ii) Sader and Jarvis [10]. Giessibl [9] performed a direct numerical discretization of Eq. (1) and thereby converted it to an equivalent matrix system. In contrast, Sader and Jarvis [10] approximated the kernel (defined below) of Eq. (1) and in so doing obtained an explicit analytical formula for the interaction force. Both methods for arbitrary amplitude have been shown to agree well for a range of (model) standard force laws [23].

Importantly, Eq. (1) is a first-kind integral equation—direct numerical solution of such equations is a notoriously ill-posed problem [24, 25]. That is, if Eq. (1) is solved directly then small errors in the measured input, $\Omega(z)$ and $a$, can potentially lead to large and unphysical excursions in the recovered force, $F(z)$. Because such input errors are unavoidable in practice, due to measurement uncertainty, an assessment of the ill-posed nature of Eq. (1) is essential. This is the focus of the present article, whose findings are critical to the robust analysis and execution of quantitative FM–AFM force spectroscopy measurements.

Ill-posedness often results from a blurring effect of the integral equation's kernel where information in the original signal is lost; the kernel is the function under the integration sign that specifies the convolution process with the original signal. Some common examples are in the field of optics and image processing, where the measured image is blurred relative to the true image of the original object [26, 27]. The aim in such cases is to reconstruct the true image from the measured blurred signal, i.e., the image is to be 'deblurred'. Clearly, such problems are ill-posed with small changes in the measured signal (due to noise) corrupting the reconstructed image—which often contains spurious oscillations absent from the true image. To alleviate this difficulty, a range of 'regularization' procedures have been formulated that aim to suppress the effects of noise in the inversion process [26, 27]. Such effects have not been considered in FM–AFM force spectroscopy. In this article, we show that FM force spectroscopy can be ill-posed in many practical measurements, with the shape of the force law directly controlling this property; see Fig. 1 which summarizes our principal findings.

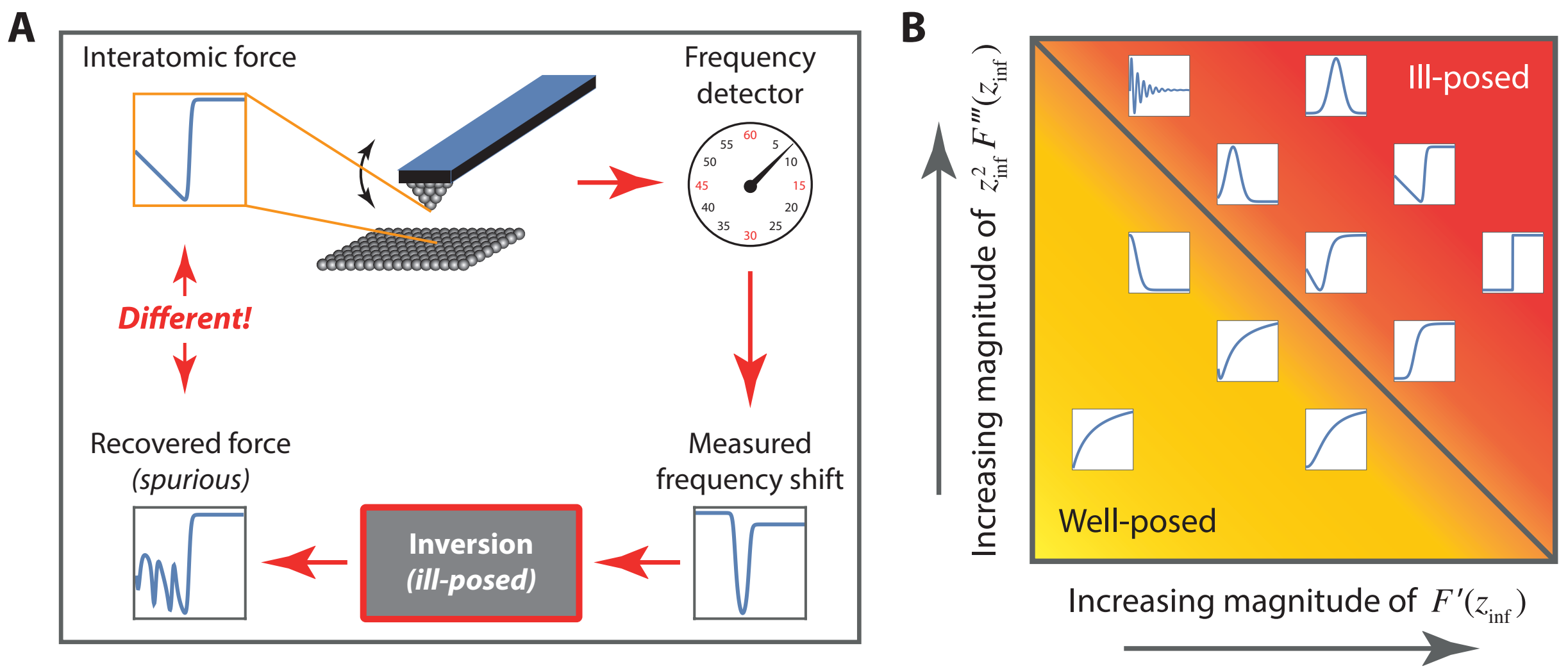

**FIG. 1: Ill-posed FM force spectroscopy and phase diagram of force law shape.** (**A**) FM force spectroscopy measurement with the recovered force differing from the true force due to ill-posedness. (**B**) Master plot of practical force laws, $F(z)$, of different shape (white boxes) and their position in the ill-posedness phase space. This position depends on the nature of the force law's inflection points, $z_{\rm inf}$. Rapidly varying force laws can produce ill-posed behavior (upper right), whereas slowly varying laws guarantee well-posed measurements (lower left). The diagonal line is specified by the 'inflection point test' (see Eq. (12)) and delineates this behavior.

To examine why Eq. (1)—which underpins FM–AFM force spectroscopy—produces such a blurring effect, and thus exhibits ill-posed behavior, we first consider a step function for the force, i.e., $F(z) = F_0 H(z_0 - z)$, where $F_0$ is the force amplitude, $H$ is the Heaviside step function and $z_0$ is the position of the step discontinuity; see Fig. S1. This is the limiting case of an attractive force, for example between two atoms, which increases rapidly on approach and has been measured using FM force spectroscopy, e.g., see Ref. [12].

It is instructive to study the equivalent form [7] of Eq. (1),

$$\Omega(z) = -\frac{1}{\pi k} \int_{-1}^{1} F'(z + a(1+u))\sqrt{1-u^2}\, du, \tag{2}$$

obtained by integrating by parts, where $F'(z)$ is the spatial derivative of the force $F(z)$. For the step function force law, we then have $F'(z) = -F_0\,\delta(z - z_0)$ where $\delta$ is the Dirac delta function. The corresponding relative frequency shift, $\Omega(z)$, which follows directly from Eq. (2), is

$$\Omega(z) = \frac{F_0}{\pi k a^2} \left[H(z - [z_0 - 2a]) - H(z - z_0)\right] \sqrt{(z_0 - z)\,(z - [z_0 - 2a])}, \tag{3}$$

and is nonzero only for $z_0 - 2a < z < z_0$; this is also plotted in Fig. S1. Comparing Eq. (3) to the original Dirac delta function for the force gradient, $F'(z)$, immediately highlights the blurring effect of Eq. (2)—and hence Eq. (1), because they are equivalent. The sharp interface at $z = z_0$ is lost and smeared onto the finite interval $z_0 - 2a < z < z_0$, as illustrated in Fig. S1. Thus, any attempt to numerically reconstruct the original force law, $F(z)$, from the corresponding relative frequency shift, $\Omega(z)$, defined in Eq. (3), is expected to be ill-posed. This is explored below.

A critical question then arises:

> *Why have previous studies (e.g., Refs. [8–19]) that directly make use of Eq. (1) apparently not shown any indication of ill-posed behavior, when applied to force laws commonly encountered in practice?*

To answer this question, we begin by expressing Eq. (1) in a more natural and equivalent form involving an explicit convolution:

$$\Omega(z) = \int_{z}^{\infty} K(t - z) F(t)\, dt, \tag{4}$$

where the kernel of this integral equation (plotted in Fig. 2A) is

$$K(x) = \frac{a - x}{\pi k a^2 \sqrt{x\,(2a - x)}} \left[H(x) - H(x - 2a)\right], \tag{5}$$

and is zero outside of the interval $0 \le x \le 2a$.

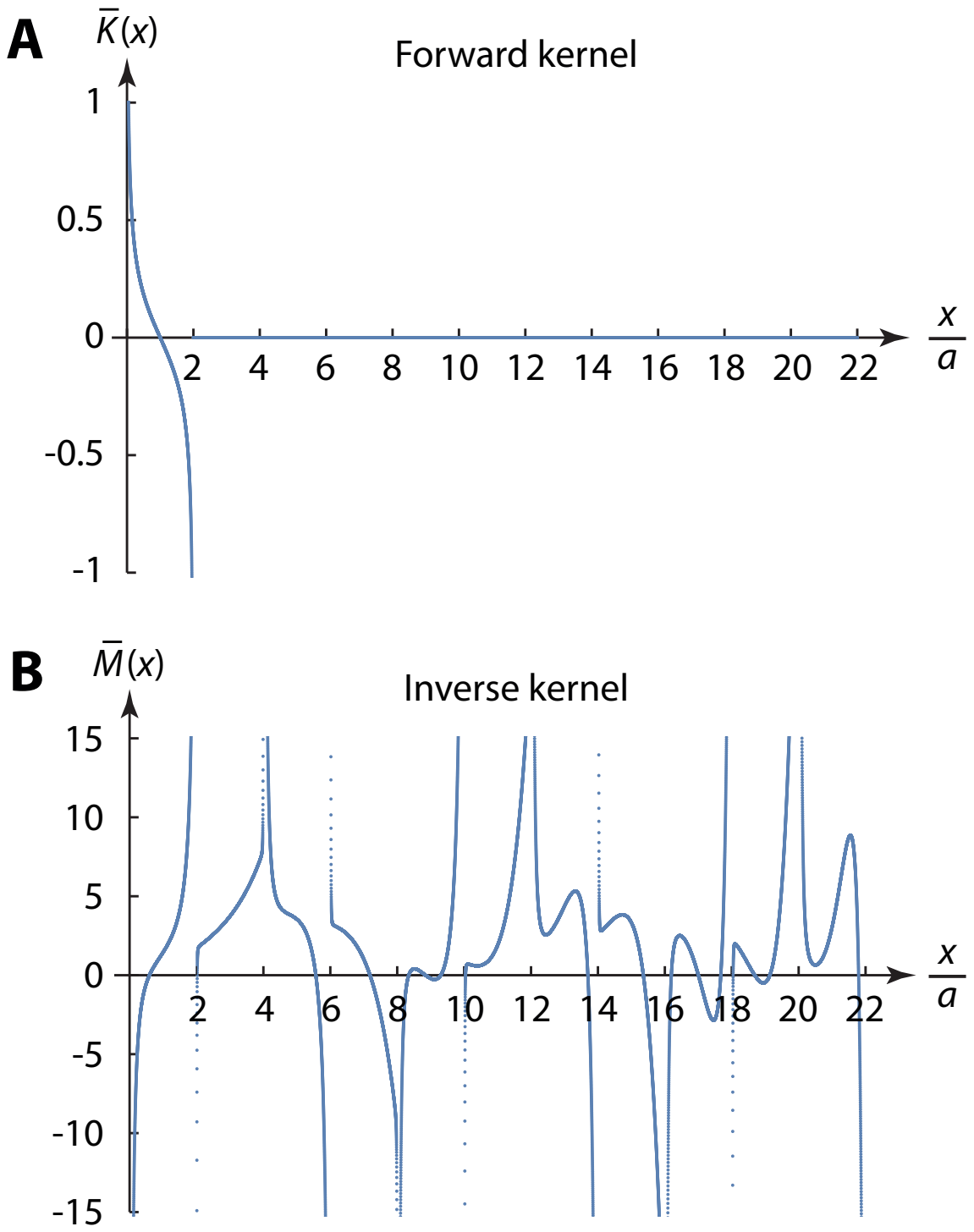

**FIG. 2: Kernels of the forward and inverse operators.** Plots showing (**A**) normalized forward kernel, $\bar{K}(x) = ka^2 K(x)$, for Eq. (4); and (**B**) the corresponding inverse kernel, $\bar{M}(x) = M(x)/(ka^2)$, for Eq. (6), computed numerically using 32,000 discrete steps. Obtained by increasing the number of steps until any differences are not discernible on the scales shown. Both kernels are zero for $x < 0$.

Equation (4) can be inverted numerically, through use of quadrature, from which we find the corresponding convolution form for its solution:

$$F(z) = \int_z^\infty M(t - z)\Omega(t)\, dt. \tag{6}$$

The associated (inverse) kernel, $M(x)$, is plotted in Fig. 2B, and takes nonzero values for arbitrarily large positive $x$ values. Results in Fig. 2B are obtained numerically on the interval $0 \leq x/a \leq 22$ using the matrix method of Giessibl [9]. All calculations are performed in Mathematica. Equation (6) establishes that the recovered force, $F(z)$, at any position, $z = z_1$, depends on the relative frequency shift, $\Omega(z)$, at that position and all larger positions, i.e., $z \geq z_1$. This feature is explicit in the method of Sader and Jarvis [10], as required.

The results in Fig. 2B show that the inverse kernel, $M(x)$, for Eq. (6) is highly oscillatory with an infinite number of singularities at $x = 2na$, where $n = 0, 1, 2, \ldots$ . Thus, a small error in the oscillation amplitude, $a$, will shift these singularities to different spatial positions. This can potentially lead to ill-posed behavior, with the recovered force, $F(z)$, varying strongly with the oscillation amplitude, $a$. Such behavior should be contingent on the functional form of the measured relative frequency shift, $\Omega(z)$.

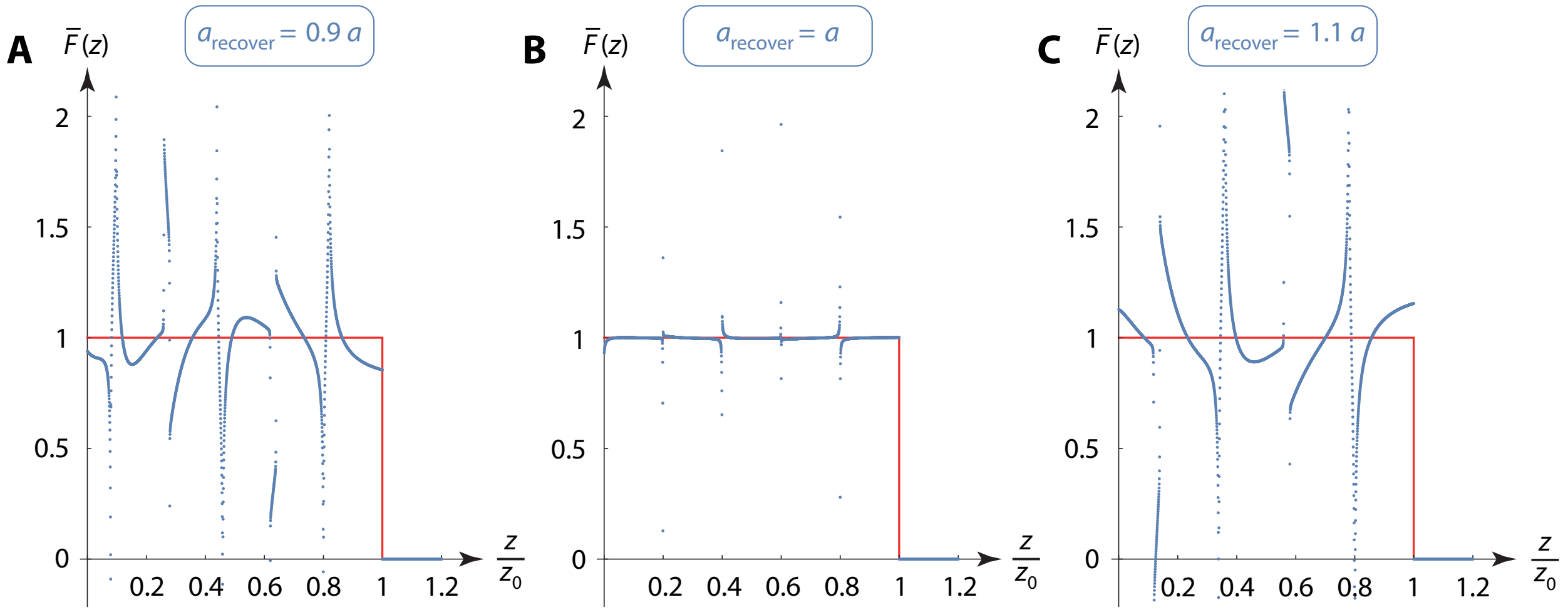

**FIG. 3: Ill-posed behavior of step force law.** Comparison of original step force (red solid straight lines) and recovered force (blue dots) using the matrix method. Normalized force, $\bar{F}(z) = F(z)/F_0$. Oscillation amplitude of $a = 0.1z_0$ is used to calculate the frequency shift, $\Omega(z)$, from which the force is recovered using the matrix method with $a_{\text{recover}} = 0.9a$ (**A**), $a$ (**B**) and $1.1a$ (**C**).

### Ill-posed behavior for a step force law

This expectation is assessed first for the step function force law. We use for this purpose the matrix method of Giessibl [9], which involves a direct numerical discretization of Eq. (1)—the same method used to generate the data in Fig. 2B (the method of Sader and Jarvis is applied later). Equation (3) for the frequency shift, $\Omega(z)$, is used as input and evaluated at 3,000 discrete points on the interval $0 \leq z \leq 1.2z_0$ with an oscillation amplitude, $a = 0.1z_0$, where $z_0$ is the position of the step discontinuity in the force, $F(z)$. The level of discretization used in measurement can be much lower, with 100 spatial points not being atypical. The matrix method is then applied to recover the force, $F(z)$, for the true oscillation amplitude, $a_{\text{recover}} = a = 0.1z_0$, and two other amplitudes, $a_{\text{recover}} = 0.9a$ and $1.1a$, corresponding to $\pm 10\%$ uncertainty—this also is not atypical in measurements. The subscript 'recover' refers to the amplitude used to recover the force.

Results for the recovered force, $F(z)$, are given in Fig. 3 and display a strong sensitivity to the chosen oscillation amplitude, $a_{\mathrm{recover}}$ (see supplementary materials for an animation showing the effect of a continuous change in $a_{\mathrm{recover}}$). Strikingly, a 10% change in $a_{\mathrm{recover}}$ produces a large variation in the recovered force, with the anomalies exhibiting similar features to the inverse kernel, $M(x)$, in Fig. 2B. Even choosing the true value for the amplitude, $a_{\mathrm{recover}} = a$, used to generate data for $\Omega(z)$, leads to strong anomalies in the recovered force, $F(z)$. This is due to discretization error, since the input frequency shift and kernel have been sampled over a finite set of discrete $z$-values. Lowering the level of discretization, i.e., reducing the number of discrete spatial points to 100, enhances the error in the recovered force for $a_{\mathrm{recover}} = a$ (see Fig. S2). This strong sensitivity to error in the input is a signature of ill-posed behavior, demonstrating that force recovery in FM–AFM can be ill-posed.

Such anomalies are yet to be reported. To date, well-posed behavior has either existed or been assumed: the recovered force has apparently been insensitive to uncertainty in the oscillation amplitude and/or frequency noise (random or systematic). Importantly, solution to Eq. (1) is inherently regularized to random noise in the measured frequency shift [28, 29], unlike many other ill-posed inverse problem [24–26]. To understand why some FM force spectroscopy measurements and simulations may not be ill-posed with respect to the remaining types of uncertainty mentioned above—as previous studies have implied [8–10, 23]—we explore the mathematical properties of the governing equation, Eq. (1).

**Force laws producing well-posed behavior**

It is easy to show that exponential functions are eigenfunctions of Eq. (1) [8]. That is, if the force law, $F(z)$, is of exponential form, then the resulting frequency shift, $\Omega(z)$, will possess an identical dependence on $z$. Specifically, if $F(z) = F_0 e^{-\lambda z}$ the resulting frequency shift is $\Omega(z) = \Omega_0 e^{-\lambda z}$, where $\Omega_0 = F_0 T(\lambda a)/(ak)$, $T(x) = I_1(x)e^{-x}$ and $I_n$ is the modified Bessel function of order $n$ [30]. This shows that Eq. (1) does not modify or blur the distance dependence of $F(z)$ if it is of exponential form—in contrast to the step function; see Fig. S1. Therefore, if the force, $F(z)$, can be expressed as a linear combination of exponential functions, i.e.,

$$F(z) = \int_0^\infty A(\lambda) e^{-\lambda z} \, d\lambda, \tag{7}$$

where $A(\lambda)$ is a specified real function, the corresponding frequency shift will possess an identical form, $\Omega(z) = 1/(ak) \int_0^\infty T(\lambda a) A(\lambda) e^{-\lambda z} \, d\lambda$ [10]. Equation (7) is immediately rec-ognizable as a Laplace transform, and thus is a statement that the force, $F(z)$, can be expressed as a Laplace transform with $A(\lambda)$ being its inverse Laplace transform. Such force laws are infinitely differentiable for $z > 0$ and decay to zero as $z \to \infty$ [31].

This suggests that, provided Eq. (7) holds, the inversion problem of reconstructing the force, $F(z)$, from the frequency shift, $\Omega(z)$, should be well-posed—with Laplace space providing the regularization. Conversely, this leads to the following *criterion* for ill-posedness of the inversion problem:

> *If the inverse Laplace transform of the force, $F(z)$, does not exist, then the inversion problem of recovering the force, $F(z)$, from the frequency shift, $\Omega(z)$, may be ill-posed.*

We illustrate the value of this criterion by (i) considering the ill-posed behavior for the step force observed in Fig. 3, and (ii) re-examining conclusions from literature results regarding the robustness of force recovery using FM–AFM.

*Step force:* The step force law, $F(z) = F_0 H(z_0 - z)$, illustrated in Fig. S1 is not differentiable at $z = z_0$, and therefore its inverse Laplace transform does not exist. Our criterion predicts ill-posed behavior may occur when probing a step force using FM–AFM and such behavior is indeed observed; see Fig. 3.

Singularities in the inverse kernel, $M(z)$, are evident in the recovered step force (Fig. 3). This spurious behavior is due to 'beating' (in Eq. (6)) of the repeating singularities of $M(t - z)$—which are spaced at a distance of $2a$—with the frequency shift $\Omega(z)$, which also is of width $2a$. Any difference in the amplitude $a$ used to measure $\Omega(z)$, with the amplitude $a_{\text{recover}}$ to recover the force, breaks this synchronization of $\Omega(z)$ and $M(t - z)$. This desynchronization drives spurious behavior in the recovered force (Fig. 3) at spatial intervals of $2a$—a feature of ill-posedness in FM force spectroscopy.

*Literature studies:* Previous literature studies [8–10, 23] apparently have not shown ill-posed behavior in FM–AFM force spectroscopy. Importantly, these literature studies have theoretically examined (model) Lennard–Jones or Morse type force laws. These force laws are constructed from power-law and exponential functions in $z$, which possess inverse Laplace transforms [32], i.e., Eq. (7) holds and they belong to Laplace space. Therefore, it is not surprising that ill-posed behavior in FM–AFM force spectroscopy has not been reported in these previous studies. Spurious oscillations exhibited by the matrix method [23] for these force laws are discussed below.

### Inflection point test: a practical approach for well-posed measurements

The step function force example and the mathematical observations that have produced our criterion for possible ill-posedness reveal potential dangers for the AFM practitioner. Because experimental data for the frequency shift, $\Omega(z)$, (and force) are discrete in nature, i.e., they are generated at discrete values of $z$, the use of standard complex variable theory techniques to establish the existence of the inverse Laplace transform are of little practical value [33, 34]. Thus, the above criterion is difficult to apply to measurements. We therefore derive an alternate condition that can be easily used by practitioners to assess the ill-posedness of FM force spectroscopy measurements.

Ill-posedness in the inversion operation for the step force law is caused by a rapid (discontinuous) jump in the force. This step force can be obtained from the continuous function, $F_{\text{smooth}}(z) = F_0/(1 + [z/z_0]^n)$ in the limit $n \to \infty$, with the step jump occurring at its *inflection point*, i.e., where its curvature changes sign and $F''(z) = 0$. So long as $n \gg 1$, the continuous function, $F_{\text{smooth}}(z)$, produces similar ill-posed behavior to the step force in Fig. 3; see Fig. S3 [35]. Importantly, because the inversion operation, Eq. (6), samples measurements from finite $z\,(\geq 0)$ to infinity, spurious results in the recovered force occur only at $z$-values smaller than that of the inflection point, i.e., $z < z_{\text{inf}}$. This observation motivates a test for ill-posedness that relies on the length scale for variation of an arbitrary force law, $F(z)$, at its inflection point, $z = z_{\text{inf}}$.

For the step force law, the inversion operation blurs its discontinuous jump (length scale of zero) onto an interval of finite width $2a$ in the frequency shift, $\Omega(z)$; see Fig. S1. This shows that the oscillation amplitude, $a$, defines the minimum length scale in $\Omega(z)$. Consequently, if the length scale for a jump in an arbitrary force law, $F(z)$, is smaller than the oscillation amplitude, this information will be lost in $\Omega(z)$—leading to ill-posed behaviour. For an arbitrary force law, $F(z)$, such a length scale is given by [36]

$$L_{\text{inf}} \equiv \sqrt{\frac{-F'(z_{\text{inf}})}{F'''(z_{\text{inf}})}}, \tag{8}$$

for an inflection point at $z = z_{\text{inf}}$, where $F''(z_{\text{inf}}) = 0$ and $'$ denotes the derivative with

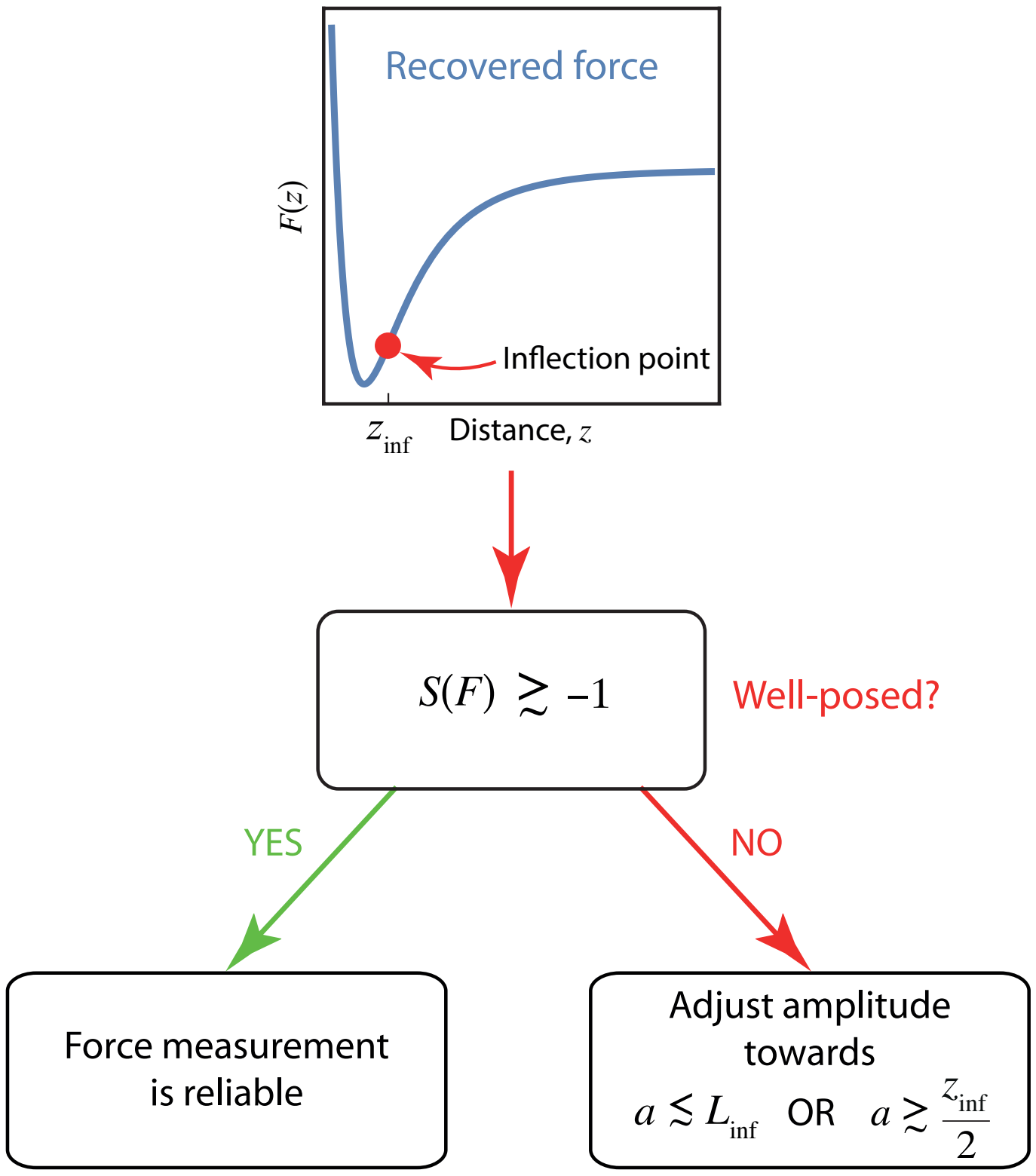

**FIG. 4: Inflection point test: assessing the ill-posedness of force measurements.** Diagram showing the required methodology to implement the key formulas in Eqs. (11) and (12). The force is first recovered from the frequency shift, to which the inflection point test is applied.

respect to $z$. From the above discussion, the first condition for ill-posed behavior is then

$$L_{\text{inf}} \lesssim a. \tag{9}$$

Inflection points in $F(z)$ generate the ill-posed behavior. Since the resulting spurious results occur in the region $z \lesssim z_{\text{inf}} - 2a$, increasing the oscillation amplitude will move these unwanted results to $z < 0$, i.e., outside the measurement range. Reversing this finding provides the second condition for which ill-posed behavior can occur:

$$a \lesssim \frac{z_{\text{inf}}}{2}. \tag{10}$$

Equations (9) and (10) are then combined to yield the required final condition for which ill-posed behavior may occur:

$$L_{\text{inf}} \lesssim a \lesssim \frac{z_{\text{inf}}}{2}, \qquad \textit{Oscillation amplitudes for ill-posedness.} \tag{11}$$

This establishes that inversion is well-posed, irrespective of the oscillation amplitude, if

$$S(F) \equiv \frac{z_{\mathrm{inf}}^2}{4} \frac{F'''(z_{\mathrm{inf}})}{F'(z_{\mathrm{inf}})} \gtrsim -1, \qquad \textit{Condition for well-posed behavior.} \tag{12}$$

Note that the amplitude of spurious behavior in the recovered force (which occurs in the region $z < z_{\mathrm{inf}}$) is comparable to the jump (increase/decrease) in $F(z)$ at the inflection point $z = z_{\mathrm{inf}}$; see Fig. 3. Inflection points that do not involve a significant jump in $F(z)$ will induce weak spurious effects. Therefore, the inflection point, $z_{\mathrm{inf}}$, used to assess ill-posedness should be selected at the largest possible $z$-value for which a strong jump in $F(z)$ occurs; see below.

This 'inflection point test' can be easily used by AFM practitioners to gauge the ill-posedness of any force measurement (via Eq. (12)), who may then adjust the chosen oscillation amplitude, $a$, to eliminate ill-posed behavior (via Eq. (11)). Its utility is evident in Fig. 1B. A flow diagram illustrating the required methodology is given in Fig. 4, which is also discussed later. Note that Eqs. (11) and (12) are to be applied to the recovered force, not the measured frequency shift. The predictive value of this test is illustrated below.

**Regularization in method of Sader and Jarvis**

Regularization typically involves replacement of the exact inverse operator, which may be unstable in the presence of noise, with an approximate but stable operator. The method of Sader and Jarvis [10] implicitly assumes that the force can be expressed in the form of Eq. (7), which enables direct inversion of the integral equation, Eq. (1). The kernel of this integral equation is approximated in the formulation of this method, with a maximum error of 5%. This approach regularizes the kernel, rendering the method insensitive to uncertainty in the oscillation amplitude; this feature is obvious from its functional form, see Eq. (9) of Ref. [10].

If the inversion problem is well-posed then the recovered force, $F(z)$, using this method will be accurate with an error commensurate with that used in approximating its kernel, i.e., 5% error. However for ill-posed measurements, any error in the formulation may still be amplified leading to spurious and unphysical results, as found for the matrix method in Fig. 3. Because uncertainty in the oscillation amplitude directly induces error in the inverse kernel, both effects can produce spurious results if the problem is ill-posed. Similarly, uncertainty in the measured frequency shift can also lead to a spurious recovered force. All these effects are explored in the results that follow.

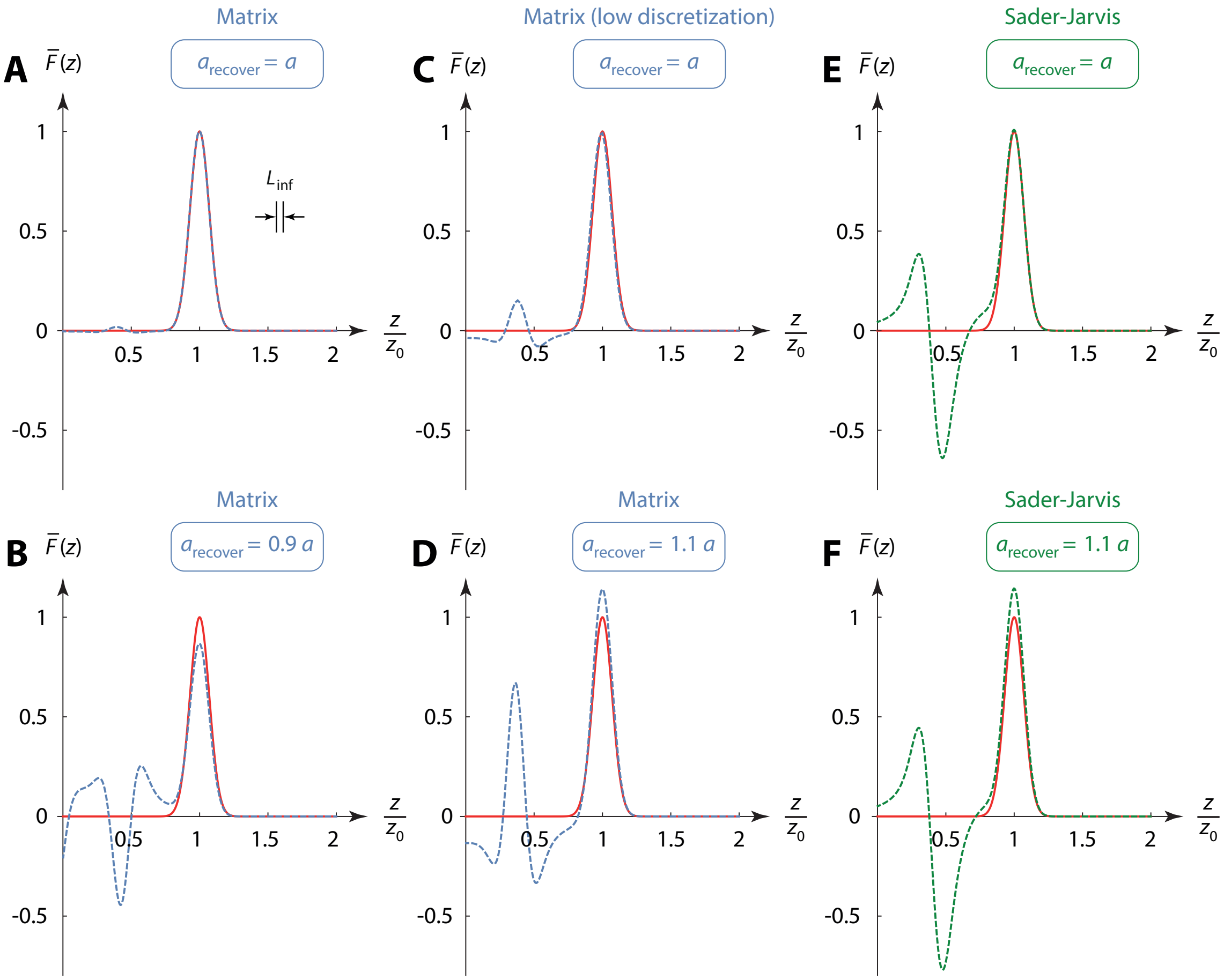

**FIG. 5: Ill-posed behavior of Gaussian force law.** Comparison of original Gaussian force (red solid curves) and recovered force (dashed curves) using the matrix method (**A**–**D**) and the Sader and Jarvis method (**E**, **F**). Normalized force, $\bar{F}(z) = F(z)/F_0$. The 'low discretization' plot (**C**) uses 100 points. Oscillation amplitude of $a = 0.3z_0$ is used to calculate the frequency shift, $\Omega(z)$, from which the force is recovered using $a_{\mathrm{recover}} = 0.9a$ (**B**), $a$ (**A**, **C**, **E**) and $1.1a$ (**D**, **F**).

### Application to a Gaussian force law

We consider a Gaussian force law, $F(z) = F_0 \exp(-[z - z_0]^2/C)$, where $F_0$ is the peak force, $C$ specifies the width of the Gaussian function and $z_0$ is the position of its peak. Similar force laws have been reported in FM force spectroscopy measurements [14] and the Gaussian peak mimics the attractive force minimum common in interatomic force laws (Fig. 1) that we experimentally investigate later. Such a force law is also highly relevant to lateral force measurements that may measure the frequency shift as a function of the spatial coordinate parallel to a surface [18]. The Gaussian force law possesses two inflection points: $z_{\rm inf} = z_0 \pm \sqrt{C/2}$, allowing evaluation of the inflection point test. Since ill-posed behavior in the recovered force is generated for $z < z_{\rm inf}$, the inflection point occurring at the larger distance, i.e., $z_{\rm inf} = z_0 + \sqrt{C/2}$, dictates whether the recovered force exhibits spurious effects within its domain, $z \geq 0$. From Eq. (11), ill-posed behaviour is predicted to occur for $\sqrt{C}/2 \lesssim a \lesssim z_0/2 + \sqrt{C}/(2\sqrt{2})$. We choose $C = z_0^2/100$ in the following investigation—for which Eq. (12) fails. Equation (11) then predicts that ill-posed behavior may be expected for

$$0.05 \lesssim \frac{a}{z_0} \lesssim 0.54. \tag{13}$$

The length scale, $L_{\rm inf} = 0.05 z_0$, over which the force jumps at its inflection point is indicated in Fig. 5A and is of order the Gaussian peak's width, as required.

Results for the Gaussian force law, that are analogous to those in Fig. 3 for the step force, are presented in Fig. 5 using 1,000 spatial points (see supplementary materials for an animation). To satisfy Eq. (13), an oscillation amplitude of $a/z_0 = 0.3$ is chosen so that ill-posed behavior may occur. Indeed, a strong sensitivity to the amplitude chosen to recover the force, when using the matrix method, is observed in Fig. 5(A–D). Strikingly, the supplementary animation shows that increasing $a_{\rm recover}$ by 1% from the true value, $a$, produces a spurious peak an order-of-magnitude larger, i.e., 10% of the true Gaussian amplitude—highlighting the hypersensitivity to amplitude uncertainty. Uncertainty in the frequency shift, $\Omega(z)$, produces similar spurious results (Fig. S4), which is discussed below. Marked spurious peaks occur at $z$-values smaller than the position of the inflection point, i.e., $z < z_{\rm inf} = 1.07 z_0$.

Note that the matrix method—which is a direct numerical solution of Eq. (1)—produces an infinite set of spurious peaks but these are mostly outside the measurement region, i.e., $z < 0$; these are explored in the next section. As for the step force, spurious peaks are also produced for the Gaussian force law by the matrix method even when there is no error in the specified oscillation amplitude. This is again due to discretization error in the kernel of Eq. (1) since a finite number of spatial points (1,000) have been chosen. Reducing this value to a discretization of 100 points, not atypical in measurements, leads to strong enhancement of the spurious peaks; see Fig. 5C.

The method of Sader and Jarvis is also found to give spurious results (Fig. 5(E, F)), but this is insensitive to the chosen amplitude—the latter property is due to regularization of the method's kernel (discussed above). These results show that the method of Sader and Jarvis is not fully regularized.

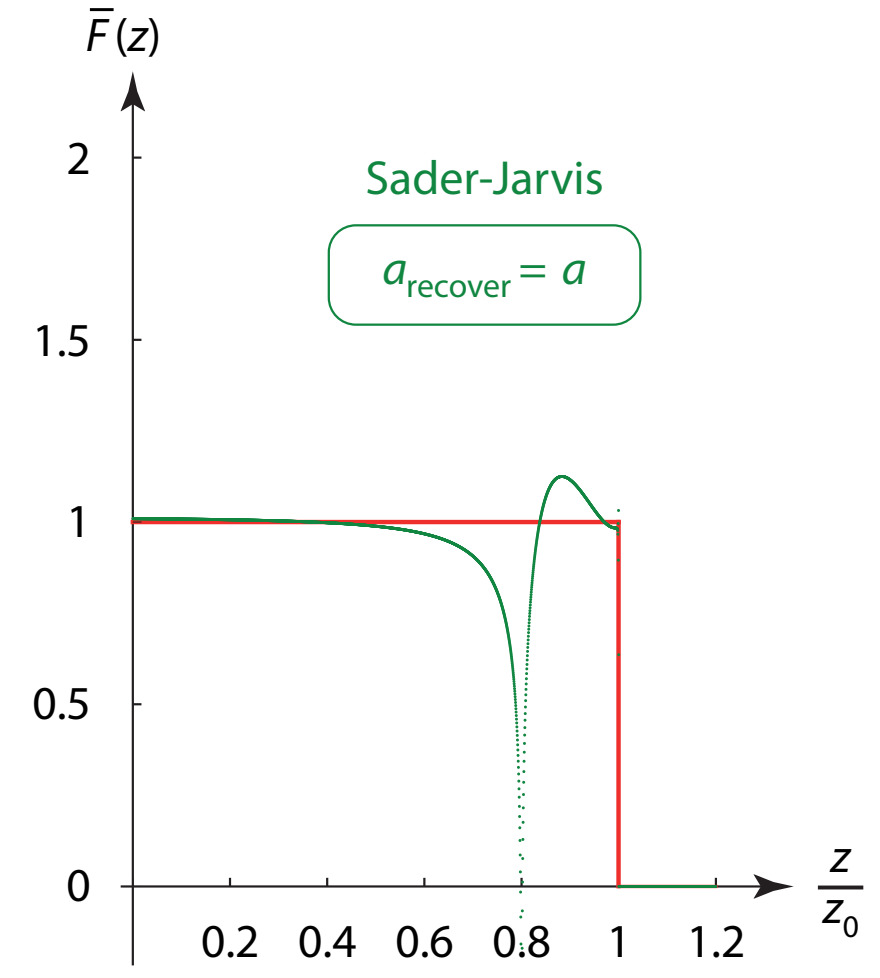

**FIG. 6: Step force law using method of Sader and Jarvis.** As for Fig. 3, but with the force recovered using the method of Sader and Jarvis (green dots) with an amplitude, $a_{\text{recover}} = a = 0.1z_0$. Similar results are obtained for $a_{\text{recover}} = 0.9a$ and $1.1a$, because the method is insensitive to small variations in $a_{\text{recover}}$.

Equations (11) and (13) indicate that increasing the oscillation amplitude used in Fig. 5 should produce well-posed behavior. This is borne out in Fig. S5 where increasing the amplitude to $a/z_0 = 1 > 0.54$ (from $a/z_0 = 0.3$) eliminates the sensitivity to oscillation amplitude and makes the inversion problem well-posed. Reducing the amplitude to $a/z_0 = 0.03$ ($< 0.05$) also gives well-posed behavior (Fig. S6), again in accord with Eq. (11). This highlights the practical utility of Eq. (11).

### Step force revisited and the effects of regularization

We now return to the step force and use the method of Sader and Jarvis [10] to recover the force from the frequency shift, under identical conditions to Fig. 3. These results are given in Fig. 6. Comparing Figs. 3 and 6 shows that the method of Sader and Jarvis also produces a spurious result, but with smoother behavior and only one spurious peak. Again, varying the oscillation amplitude has little effect on the recovered force, as required (data not shown).

For a general force law, the matrix method will exhibit an infinite number of spurious peaks in the recovered force at spatial intervals of $2a$, i.e., $z \approx z_{\rm inf} - 2na$ where $n = 1, 2, 3, \ldots$; those in the measurement region, $z \geq 0$, are visible. In contrast, the method of Sader and Jarvis will only produce one such peak at $z \approx z_{\rm inf} - 2a$. This difference is due to the inherent regularization of the inverse kernel in the Sader and Jarvis method (see above). Because the matrix method is based on a direct numerical discretization of Eq. (1), it provides no regularization [24]. These properties of the matrix method and the method of Sader and Jarvis are illustrated in Fig. S7 for a Gaussian force law with a narrower full-width-half-maximum (FWHM) than in Fig. 5.

This study also sheds light on spurious oscillations in the matrix method that are observed to grow as the amplitude is decreased (see Ref. [23]); these oscillations are sensitive to small changes in amplitude. In the limit of small amplitude, the force is simply the integral of the frequency shift [20]. The matrix method recovers this property using (a discrete version of) Eq. (6) involving the highly oscillatory/singular inverse kernel (Fig. 2B). This demands balance and cancellation of the kernel's singularities. Obviously, reducing $a$ enhances discretization error in the kernel, obviating this requirement, leading to spurious effects—even though the model force laws studied [23] belong to Laplace space. The latter provides a natural regularization that reduces the effects of any ill-posedness. These spurious oscillations are absent in the method of Sader and Jarvis again due to regularization of its kernel.

**Practical considerations**

These findings highlight the importance of avoiding ill-posed behavior in practical FM force spectroscopy. Otherwise, spurious and unphysical force measurements can result. Fortunately, Eq. (11) shows that this is immediately possible by simply tuning the oscillation amplitude appropriately. If the primary condition in Eq. (12) is violated, the oscillation amplitude, $a$, should be increased or decreased systematically in the frequency shift measurement and recovered force, according to Eq. (11)—until the recovered force is independent of the chosen oscillation amplitude, to within measurement uncertainty; see Fig. 4. Equation (11) can also be used independently because Eq. (12) is derived from it.

Uncertainty always exists in measurements of the frequency shift and oscillation ampli-tude, discretization error occurs because the frequency shift is sampled at a finite number of spatial positions, and the precise form of the interaction force in a measurement is normally unknown. Furthermore, a minor measurement anomaly in $\Omega(z)$ at a localized position $z$, e.g., caused by a rapid and unanticipated change in local curvature from a $z$-piezo error, can cause its inverse Laplace transform to be non-existent, rendering the inversion process ill-posed. This could result in spurious behavior in the recovered force, $F(z)$, for smaller values of $z$—similar to that observed for the (model) step and Gaussian force laws. Therefore, elimination of spurious force measurements arising from the ill-posed nature of the inversion process, by simply adjusting the oscillation amplitude used in the force recovery method to its true value (i.e., setting $a_{\text{recover}} = a$, as in Figs. 3 and 5), is generally not possible in practice and should not be attempted. An example of such spurious behavior is given in Fig. S4 for a Gaussian force law that contains a minor (1%) anomaly in the $z$-dependence of its frequency shift. Force recovery is implemented using the true amplitude, i.e., $a_{\text{recover}} = a$. While the frequency shift anomaly is not visible (Fig. S4B), it has a striking effect on the recovered force using the matrix method, while the method of Sader and Jarvis produces the same spurious peak as in Fig. 5(E, F) and is insensitive to this anomaly.

Sensitivity to $a_{\text{recover}}$ using the matrix method indicates ill-posedness. However, it is possible for the inversion problem to be ill-posed but display little variation with $a_{\text{recover}}$. Such a sensitivity analysis should therefore not be used to assess the well-posedness of measurements. The only robust approach available to ensure well-posed behavior is to adjust the amplitude in individual measurements, in accordance with Eq. (11), until the measured force is insensitive to the amplitude used (discussed above).

**An ill-posed atomically-resolved measurement**

The above theoretical findings are now assessed experimentally. Figure 7 shows atomically-resolved measurements of a single atom Cu tip over a Cu adatom on a Cu(111) surface with a low-temperature (5.9 K) AFM in ultrahigh vacuum (supplementary materials). While the measured frequency shift looks similar to Lennard-Jones or Morse type laws, the measurement is clearly ill-posed. The recovered force from the matrix method is highly sensitive to the chosen oscillation amplitude and results of both methods in Fig. 7(B, C) are strikingly different. Indeed, the matrix method produces 5× variation in the force gradient

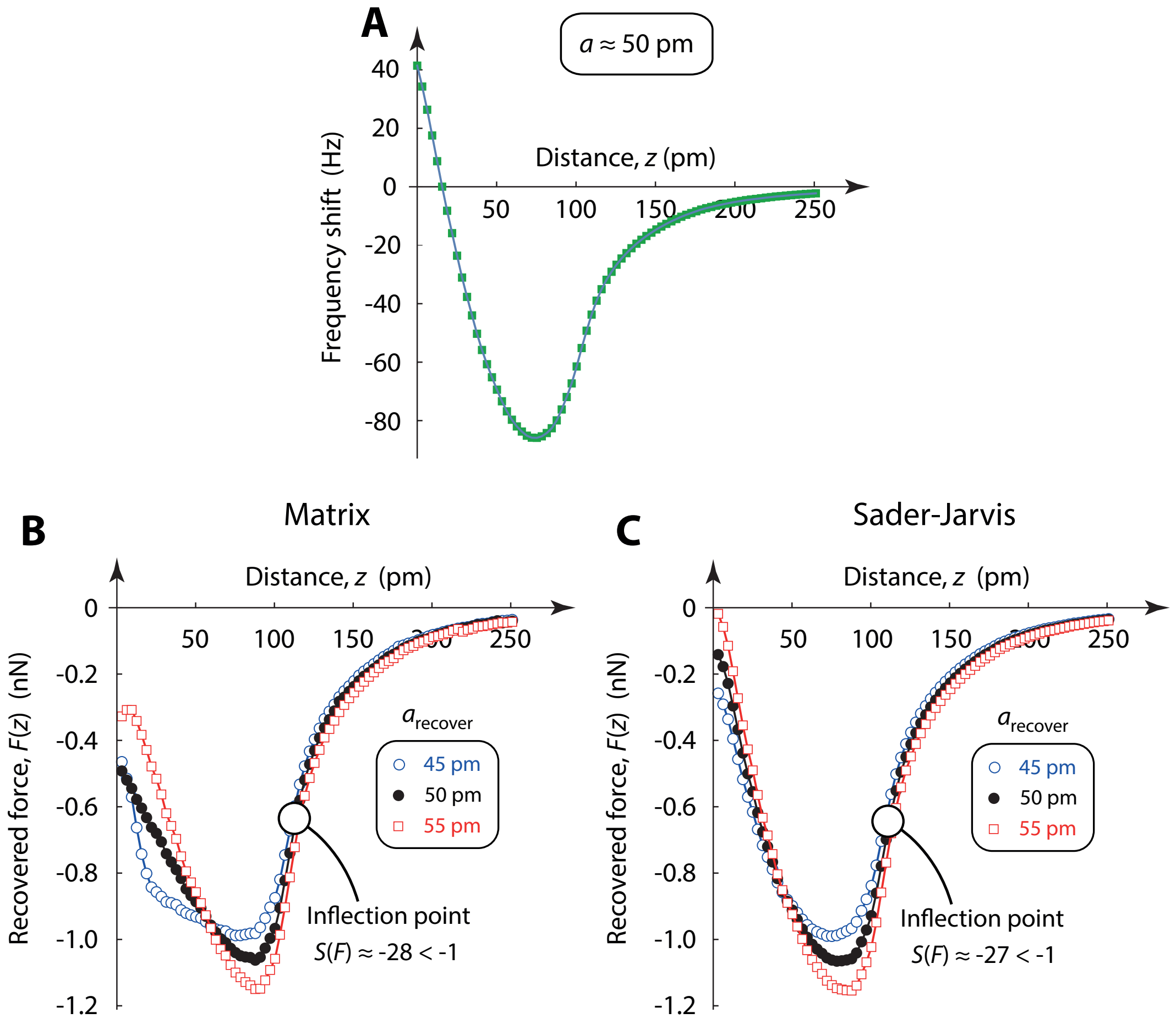

**FIG. 7: Ill-posed atomically-resolved force measurement.** Atomically-resolved measurement of a single atom Cu tip over a Cu adatom on a Cu(111) surface at 5.9 K in ultrahigh vacuum. (**A**) Measured frequency shift, and recovered force using (**B**) matrix method, (**C**) Sader and Jarvis method. Measured oscillation amplitude is approximately 50 pm.

at $z = 50\,\text{pm}$ for $\pm 10\%$ change in $a_{\text{recover}}$.

We now demonstrate the utility of the inflection point test. The inflection point (in the recovered force) used to assessed ill-posedness is shown in Fig. 7(B, C). This yields $S(F) \approx -27$ using both the matrix method and the Sader and Jarvis method; $S(F)$ is also weakly dependent on the chosen oscillation amplitude. This value violates Eq. (12), establishing potential ill-posedness. Following Fig. 4, Eq. (11) then yields, $10\,\text{pm} \lesssim a \lesssim 54\,\text{pm}$, confirming that the measurement can indeed be ill-posed for the amplitude used: $a \approx 50\,\text{pm}$.

The results in Fig. 7(B, C) show the unreliability of this ill-posed force measurement in the region $0 < z \lesssim z_{\rm inf} - a$. Importantly, validity of the matrix method cannot be assessed by calculating the frequency shift from the recovered force, and comparing that result to the (true) measured frequency shift. This will always yield identical results because it simply reverses the original matrix solution. Also, the frequency shift determined using the recovered force from the Sader and Jarvis method will be similar to that measured—with an error ($\approx 5\%$) commensurate with the approximation of its kernel (this forward operation is well-posed). Neither the matrix method nor the Sader and Jarvis method should be used in the region $0 < z \lesssim z_{\rm inf} - a$ when ill-posedness exists—they are both unreliable in such cases.

In contrast, Fig. S8 shows a separate atomically-resolved measurement with an oscillation amplitude now chosen outside Eq. (11): $a \approx 50\,\text{pm} > z_{\rm inf}/2 \approx 25\,\text{pm}$. As expected, this yields well-posed behavior with similar results for the recovered force from the matrix method and the Sader and Jarvis method; see Fig. S8. The method of Sader and Jarvis exhibits superior noise performance due to regularization of its kernel; the matrix method has no regularization. This further illustrates the importance of regularization.

The aim of this article is to report on the ill-posedness of atomically-resolved force spectroscopy, a critical feature of inverse problems that has gone unnoticed in AFM measurements. Force laws producing this behavior are not uncommon. Ill-posed behavior is induced by a rapid change in the measured force law's concavity, rather than random frequency noise; systematic error in the measured frequency shift can produce ill-posedness. When ill-posed, small uncertainty in the measured oscillation amplitude, frequency shift and/or the discrete nature of the measured frequency shift can lead to a spurious recovered force. If the force law belongs to Laplace space this provides a natural regularization of the inverse problem, minimizing the effects of ill-posedness. A simple but effective 'inflection point test' has been formulated that guides the AFM practitioner to well-posed measurements.

Much work remains to be done, particularly in formulating appropriate regularization schemes for atomically-resolved force measurements that eliminate the effects of ill-posedness. This would in principle enable measurements at all oscillation amplitudes, regardless of the nature of the measured force law, which is highly desirable from a practical viewpoint.

The authors gratefully acknowledge support from the Australian Research Council Centre of Excellence in Exciton Science, the Australian Research Council Grants Scheme and Deutsche Forschungsgemeinschaft SFB 689/1277.

* Corresponding author. Email: jsader@unimelb.edu.au

([33], p. 213), are unnecessarily restrictive in scope and hard to apply to discretely-sampled numerical data when $n$ is large.

[35] $F_{\mathrm{smooth}}(z)$ possesses an inverse Laplace transform for arbitrary finite $n$. For $n \gg 1$, such rapidly varying functions are on the edge of Laplace space because they are discontinuous in the limit $n \to \infty$. While Laplace space regularizes ill-posedness, reducing its effects, a rapid jump in an arbitrary force law, $F(z)$, at an inflection point can still drive ill-posed behavior.

[36] Geometrically, the ratio of the slope of $F(z)$ to the rate-of-change in its curvature gives the square of the required length scale at its inflection point. A jump in $F(z)$ does not occur if $F'(z_{\mathrm{inf}})$ and $F'''(z_{\mathrm{inf}})$ are of identical sign or $F'(z_{\mathrm{inf}}) = 0$, and as such, these cases are not relevant here. Equation (8) is also undefined if $F'''(z_{\mathrm{inf}}) = 0$ and hence $F^{(\mathrm{iv})}(z_{\mathrm{inf}}) = 0$, but this is an unlikely practical scenario.

# SUPPLEMENTARY MATERIALS

## Interatomic force laws that corrupt their own measurement

John E. Sader[1,*], Barry D. Hughes[1], Ferdinand Huber[2], and Franz J. Giessibl[2]

[1] *School of Mathematics and Statistics, The University of Melbourne, Victoria 3010, Australia*

[2] *Institute of Experimental and Applied Physics, University of Regensburg, D-93053 Regensburg, Germany*

[*]Corresponding author. Email: jsader@unimelb.edu.au

# Methods for atomically-resolved FM force spectroscopy measurements

All experiments were performed with a custom-built low temperature scanning probe microscope operating at 5.9 K in ultra-high vacuum. We used a qPlus sensor [1] which was equipped with an etched tungsten tip and had a stiffness of 1,800 N/m, a resonance frequency of 29.191 kHz and a quality factor of 31,788. The sensor was operated in frequency-modulation mode [2] at an amplitude of 50 pm.

The Cu(111) sample was cleaned by repeated sputtering and annealing cycles. Less than 0.01 monolayer of CO was adsorbed on the surface by dosing CO into the microscope's ultra-high vacuum chamber. Single Cu adatoms were deposited by using a custom-built Cu evaporator consisting of a small piece of a Cu wire wrapped in a tungsten filament. Both depositions were done while the sample was in the microscope and cooled down to the microscope's base temperature of 5.9 K.

A sharp, single monoatomic tip was created after multiple indentations of the tip into the clean Cu surface and subsequent identification by the method described in Refs. [3, 4]. A CO tip was created by picking up a CO from the surface with a recipe similar to the one in Ref. [5]. Two frequency shift-distance spectra above the center of the CO (or the Cu adatom) and the clean Cu surface in the same distance range were recorded and subtracted from each other to obtain the short-range frequency shift and the force subsequently [6].

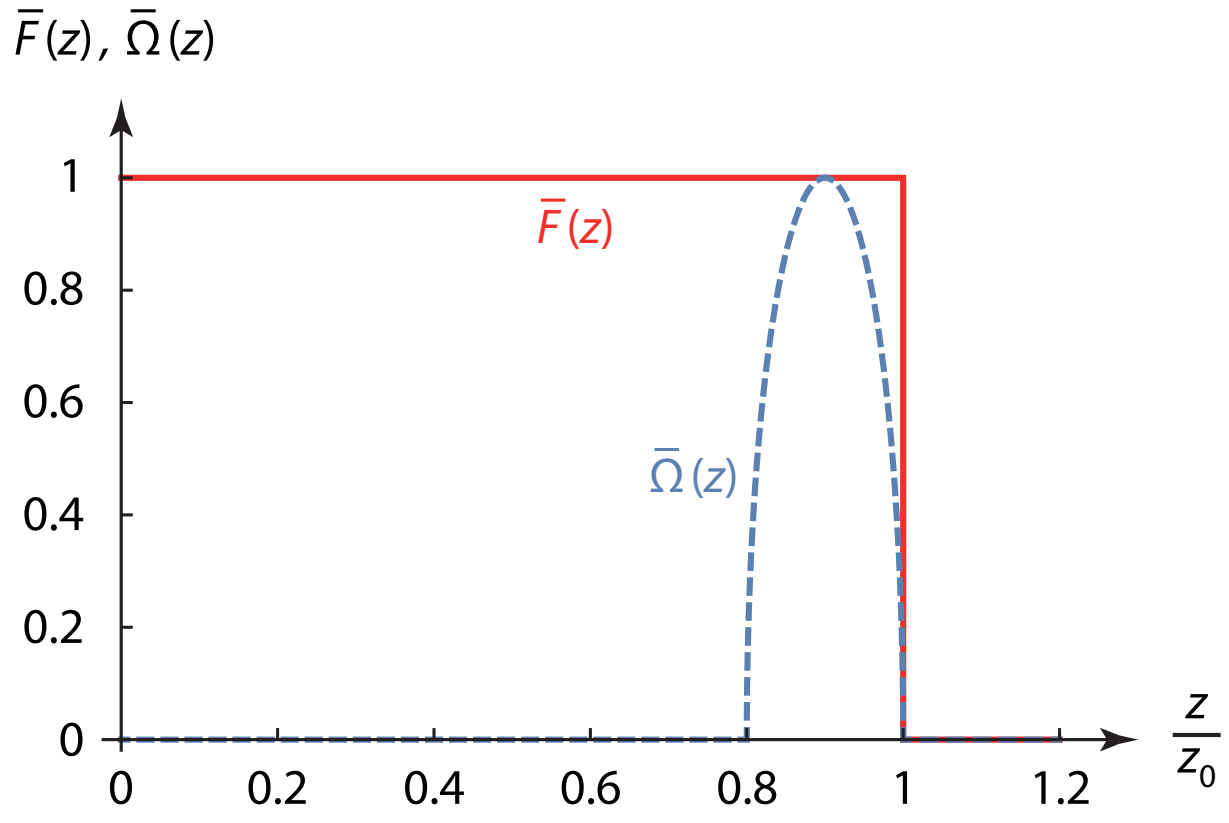

**FIG. S1**: **The step force law.** Plot showing normalized step function force law, $\bar{F}(z) = F(z)/F_0$ (red solid curve) and its corresponding relative frequency shift, $\bar{\Omega}(z) = \Omega(z)\pi k a/F_0$ (blue dashed curve), in Eq. (3). Results are given for $a = 0.1\, z_0$.

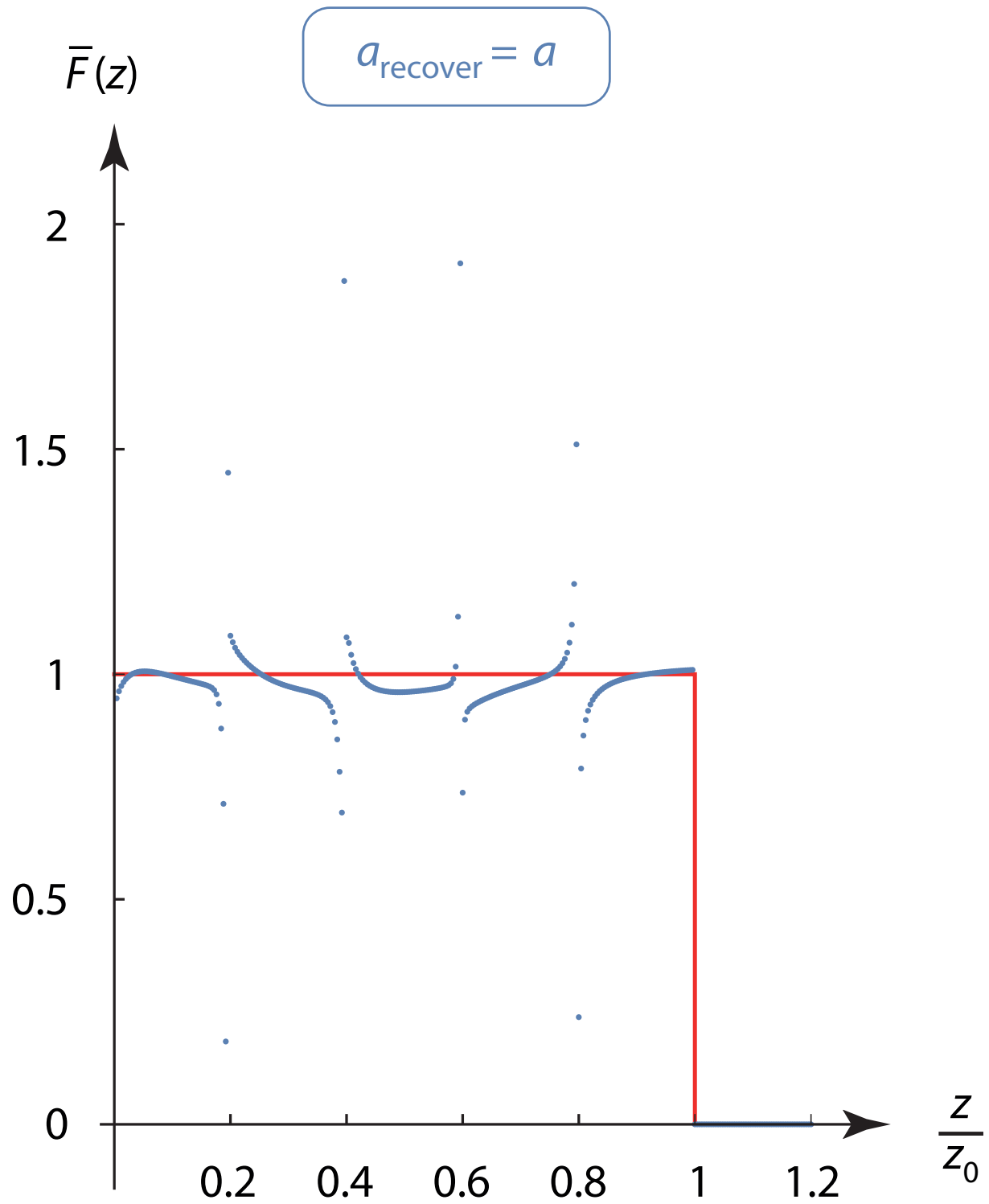

**FIG. S2**: **Ill-posed behavior of step force law with lower spatial discretization.** As for Fig. 3B ($a = a_{\text{recover}}$) but with a lower discretization of 300 spatial points in the matrix method (instead of 3,000 points). Comparing this result to Fig. 3B shows that a lower discretization enhances the discrepancy between the true and recovered force. This is due to ill-posedness of the inversion problem and greater error in the (discrete) representation of the inverse kernel. Results are given for $a = 0.1\, z_0$.

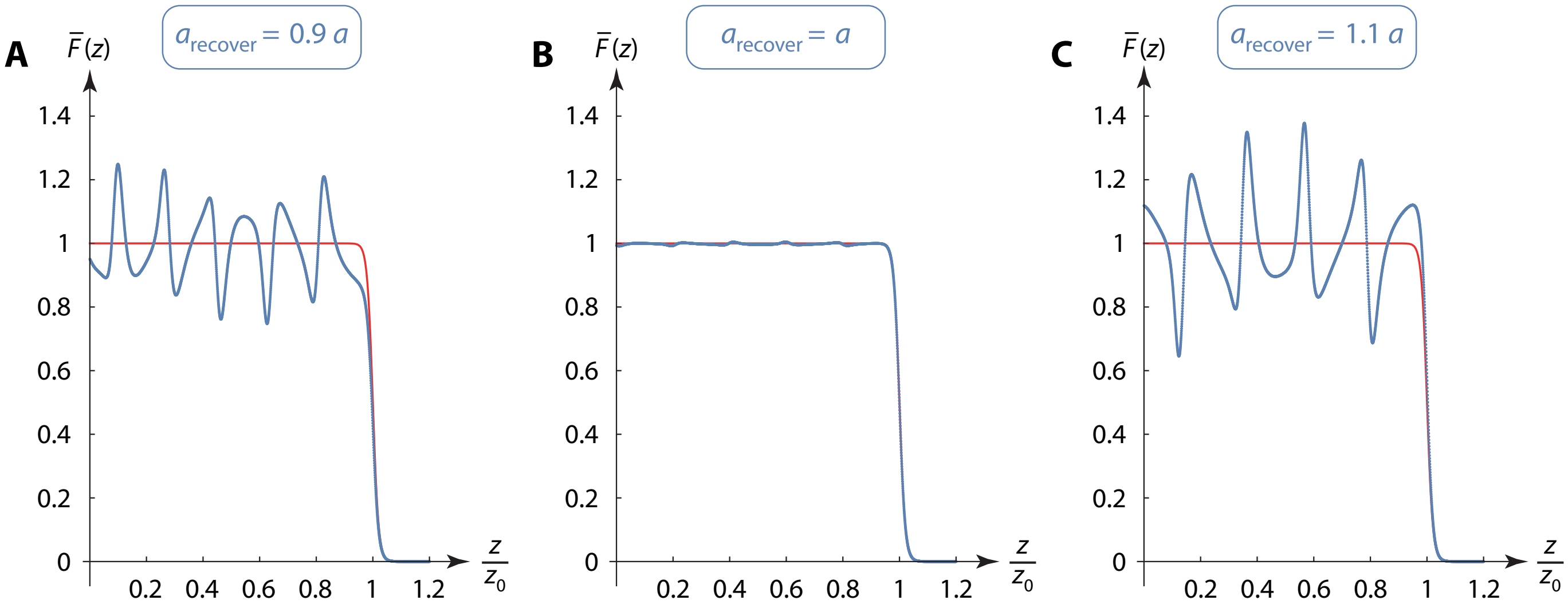

**FIG. S3**: **Ill-posed behavior of a smoothed step force law.** Comparison of original smoothed step force $F_{\text{smooth}}(z) = F_0/(1 + [z/z_0]^{100})$ (red solid lines) and recovered force (blue dots) using the matrix method with 3,000 spatial points. Normalized force, $\bar{F}(z) = F_{\text{smooth}}(z)/F_0$. Oscillation amplitude of $a = 0.1z_0$ is used to calculate the frequency shift, $\Omega(z)$, from which the force is recovered using the matrix method with $a_{\text{recover}} = 0.9a$ (**A**), $a$ (**B**) and $1.1a$ (**C**); the subscript 'recover' refers to the amplitude used to recover the force. Comparing these results to those in Fig. 3 shows that a force law exhibiting a continuous rapid change in concavity can produce ill-posed behavior (spurious effects).

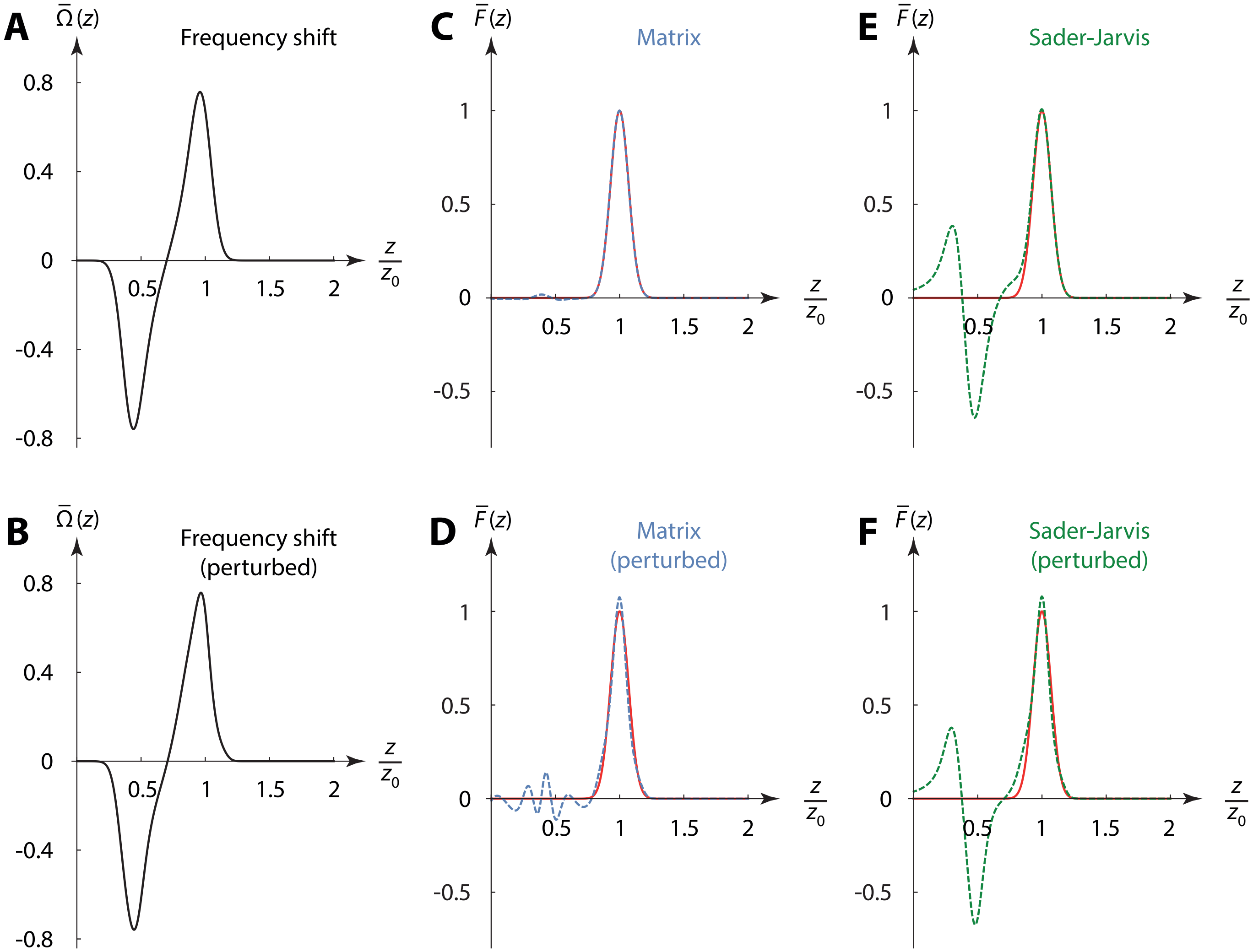

**FIG. S4: Ill-posed behavior of Gaussian force law with $a_{\text{recover}} = a$ and a small perturbation to the frequency shift.** Comparison of frequency shift (**A**, **B**) and recovered force (**C**–**F**) for an original Gaussian force, $F(z) = F_0 \exp(-[z - z_0]^2/C)$, with $C = z_0^2/1000$ (red solid curves) and recovered force (dashed curves) using (**C**, **D**) the matrix method, and (**E**, **F**) the Sader and Jarvis method. Force recovery performed with $a_{\text{recover}} = a = 0.3z_0$, i.e., no amplitude uncertainty. Normalized force, $\bar{F}(z) = F(z)/F_0$ and normalized frequency shift, $\bar{\Omega}(z) = \Omega(z)\pi k a/F_0$, are shown. A discretization of 1,000 steps on the interval $z \in [0, 2z_0]$ is used. (**A**, **C**, **E**) No perturbation in the frequency shift; identical to Fig. 5(A, E, F). (**B**, **D**, **F**) Small perturbation introduced where the $z$-value used to calculate the frequency shift (from the original force) is chirped by 1% using the transformation $z \to z(1+\frac{1}{100}\sin(8\pi z))$—simulating a $z$-piezo error. This perturbation is not obvious in (**B**) but has a dramatic effect on the recovered force using the matrix method with spurious oscillations one order-of-magnitude larger appearing (**D**). The frequency perturbation does not strongly affect the Sader and Jarvis method (**E**, **F**) due to its (partial) regularization—the recovered forces still deviate significantly from the true Gaussian force by similar amounts (as expected). This demonstrates that using the true oscillation amplitude can lead to a spurious recovered force when ill-posedness exist, because the inversion operation is sensitive to uncertainty in both the oscillation amplitude and the measured frequency shift.

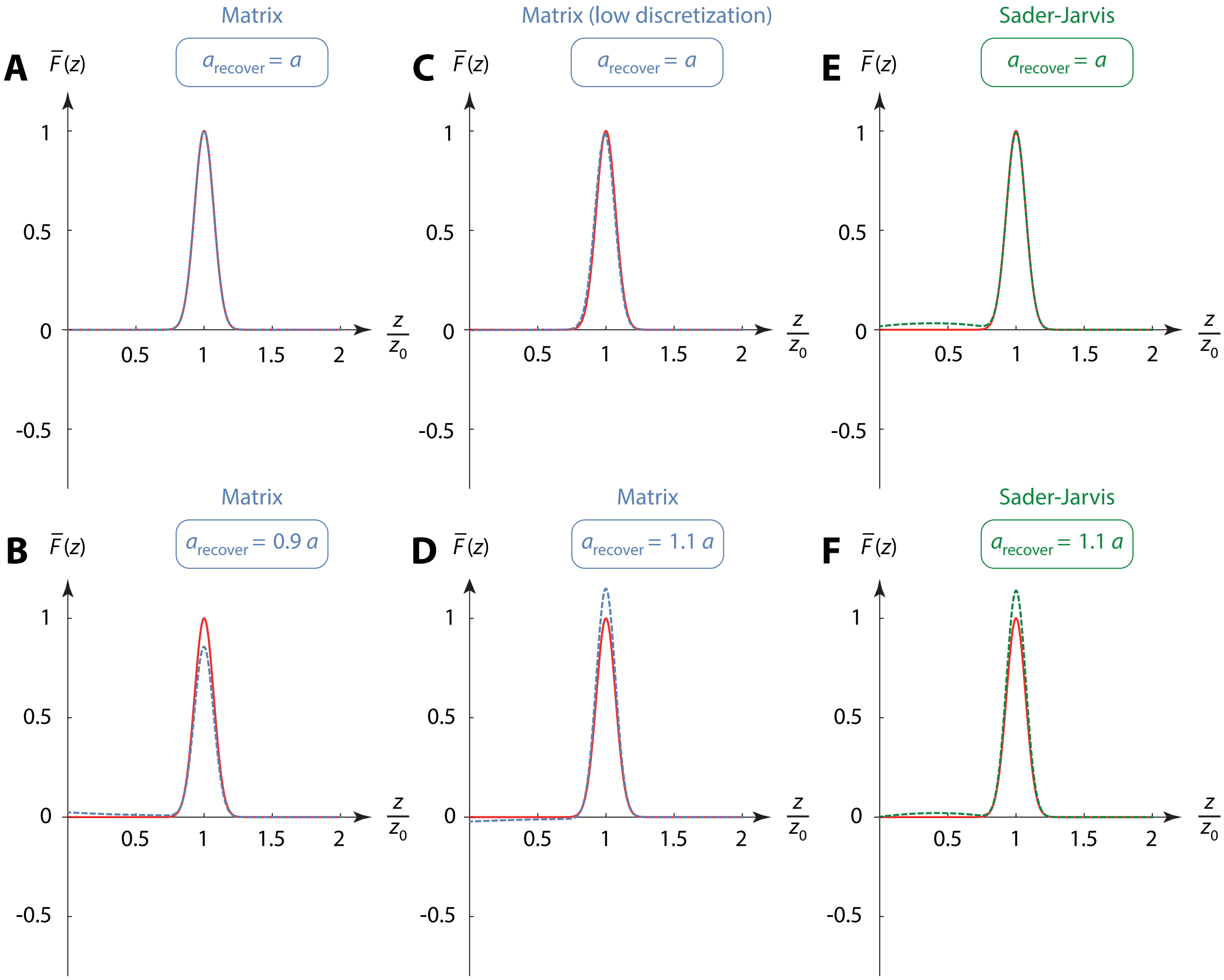

**FIG. S5**: **Well-posed behavior of Gaussian force law at larger oscillation amplitude, $a \gtrsim z_{\mathrm{inf}}/2$.** As for Fig. 5, but using an oscillation amplitude of $a = z_0$ to calculate the relative frequency shift, $\Omega(z)$, from which the force is recovered using three different amplitudes, $a_{\mathrm{recover}} = 0.9a$ (**B**), $a$ (**A**, **C**, **E**) and $1.1a$ (**D**, **F**), corresponding to $\pm 10\%$ uncertainty. Solid (red) curves are the true Gaussian force, whereas dashed curves correspond to the recovered force. 1,000 spatial points are used on the $z$-interval shown, apart from 'low discretization' (**C**) where 100 points are used. Well-posed behavior is evident. Error in the method of Sader and Jarvis (**E**, **F**) is commensurate with the 5% error in its approximate kernel.

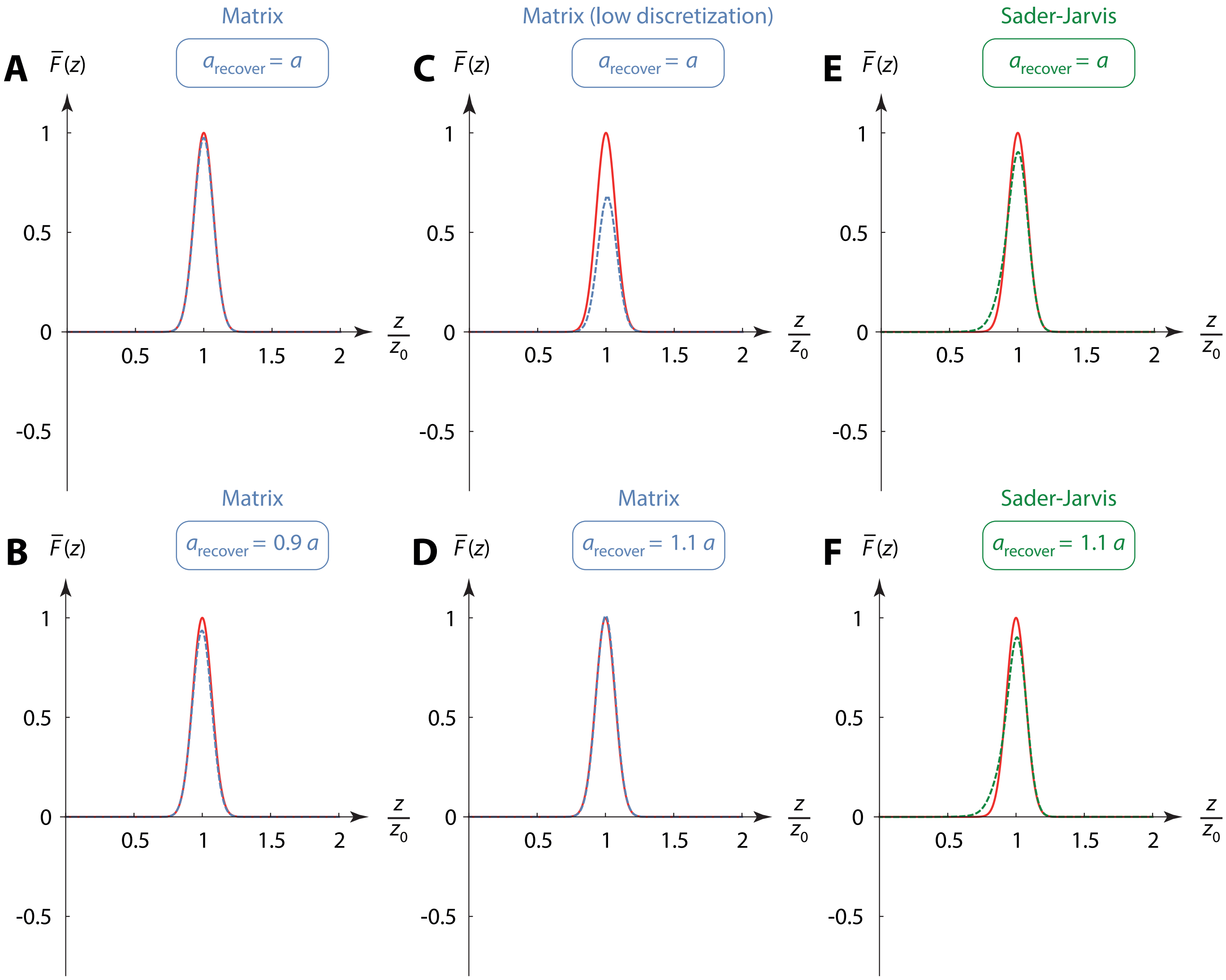

**FIG. S6**: **Well-posed behavior of Gaussian force law at smaller oscillation amplitude, $a \lesssim L_{inf}$.** As for Fig. 5, but using an oscillation amplitude of $a = 0.03z_0$ to calculate the relative frequency shift, $\Omega(z)$, from which the force is recovered using three different amplitudes, $a_{recover} = 0.9a$ (**B**), $a$ (**A**, **C**, **E**) and $1.1a$ (**D**, **F**), corresponding to $\pm 10\%$ uncertainty. Solid (red) curves are the true Gaussian force, whereas dashed curves correspond to the recovered force. 1,000 spatial points are used on the $z$-interval shown, apart from 'low discretization' (**C**) where 100 points are used. Well-posed behavior is evident. Error in the method of Sader and Jarvis (**E**, **F**) is commensurate with the 5% error in its approximate kernel. Significant error in the matrix method for low discretization (**C**) is due to inadequate numerical representation of integrals when using a small number of spatial points (100 points) and small oscillation amplitude.

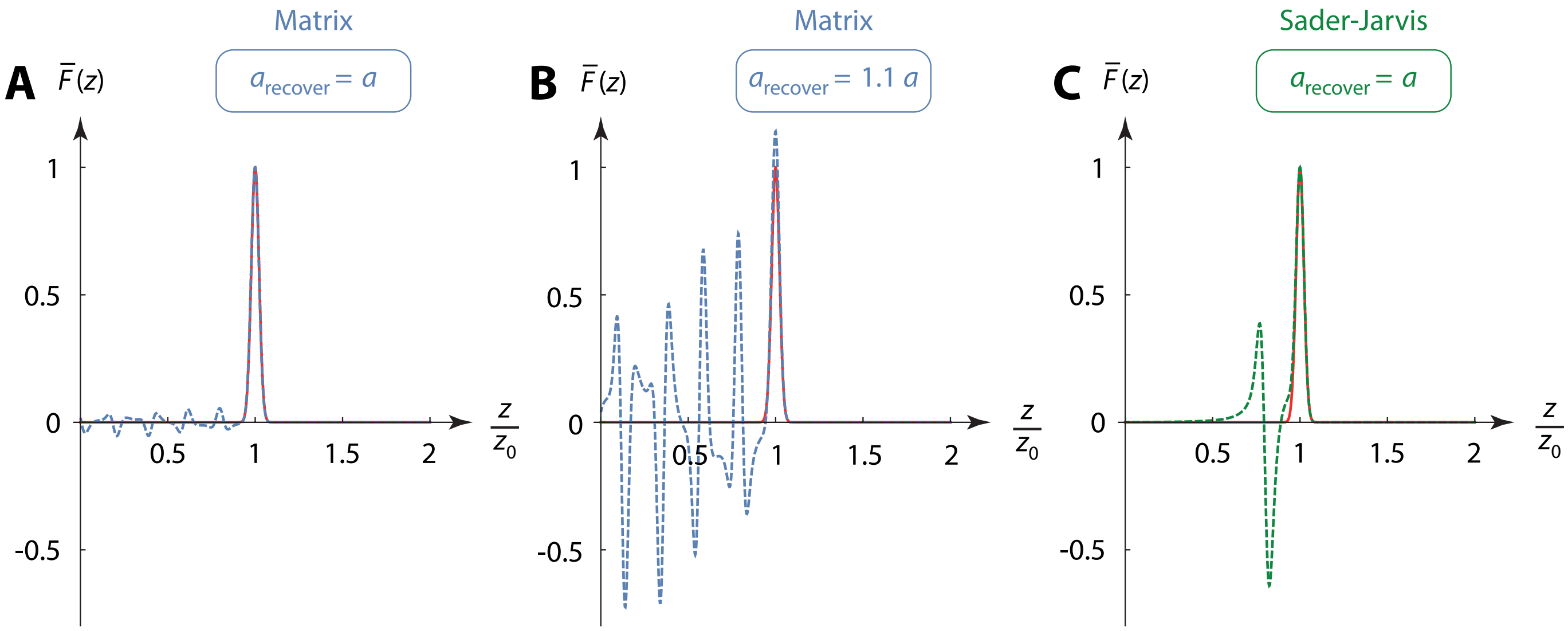

**FIG. S7**: **Ill-posed behavior of narrower Gaussian force law.** Comparison of original Gaussian force, $F(z) = F_0 \exp(-[z - z_0]^2/C)$, with $C = z_0^2/1000$ (red solid curves) and recovered force (dashed curves) using the matrix method (**A**, **B**) and the Sader and Jarvis method (**C**). A normalized force, $\bar{F}(z) = F(z)/F_0$ is shown. A discretization of 1,000 steps on the interval $z \in [0, 2z_0]$ is used. An oscillation amplitude of $a = 0.1z_0$ is used to calculate the relative frequency shift, $\Omega(z)$, from which the force is recovered using two different amplitudes, $a_{\text{recover}} = a$ (**A**, **C**) and $1.1a$ (**B**). The matrix method (**A**, **B**) shows multiple spurious peaks and is highly sensitive to $a_{\text{recover}}$; it also generates an infinite set of spurious peaks outside the measurement region, i.e., $z < 0$. The method of Sader and Jarvis (**C**) produces only one spurious peak and is insensitive to $a_{\text{recover}}$, due to its regularization.

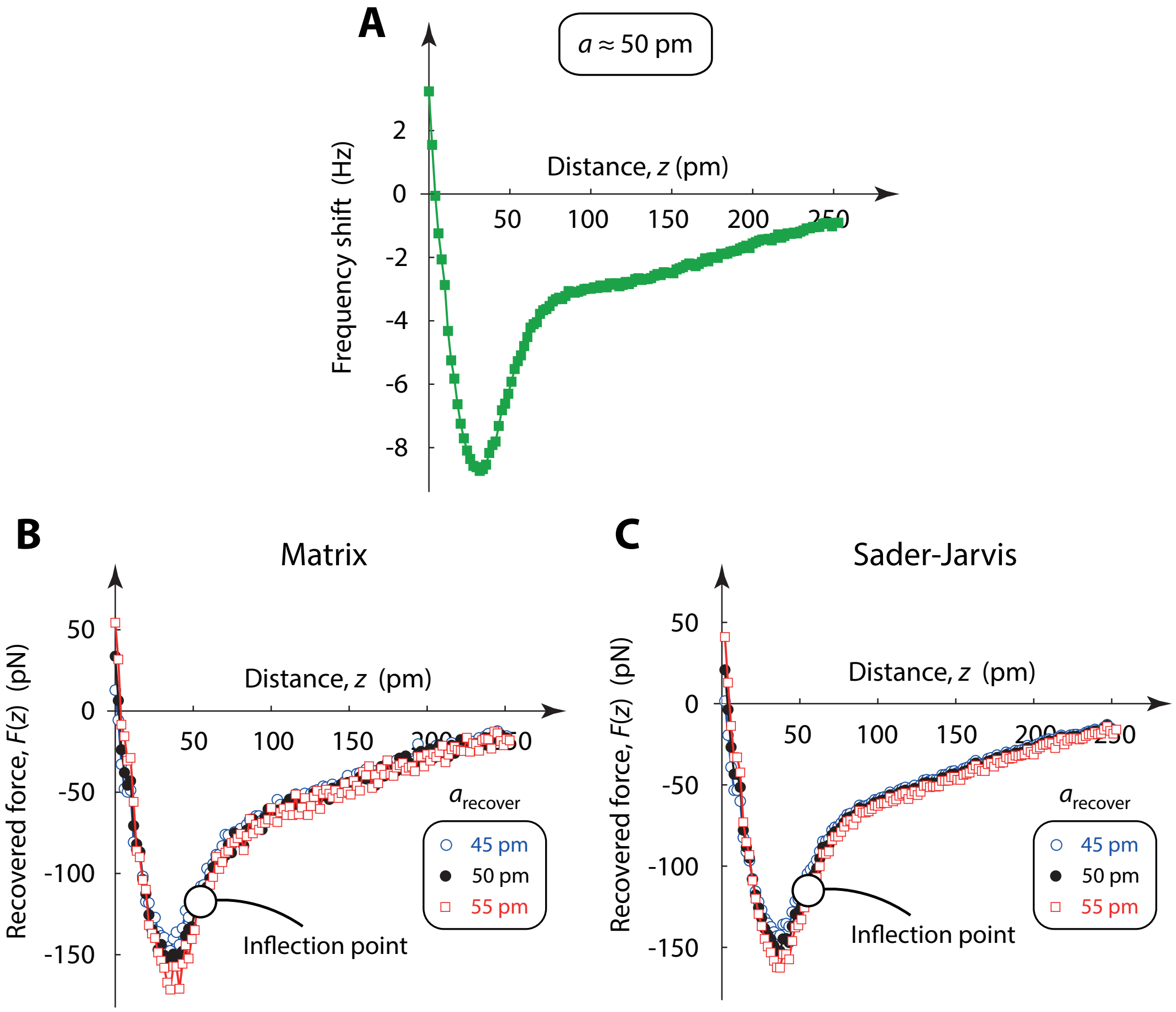

**FIG. S8**: **Well-posed atomically-resolved force measurement.** Atomically-resolved measurement of a single atom Cu tip over a CO molecule on a Cu(111) surface at 5.9 K in ultrahigh vacuum. (**A**) Measured frequency shift, and recovered force using (**B**) matrix method, (**C**) Sader and Jarvis method. Oscillation amplitude $a \approx 50\,\mathrm{pm} > z_{\mathrm{inf}}/2 \approx 25\,\mathrm{pm}$, establishing well-posedness via Eq. (11). See above for methods.